\documentclass[12pt]{article}

\usepackage{booktabs}
\usepackage{caption}
\usepackage{threeparttable}
\usepackage{subfigure}
\usepackage{amsmath}
\usepackage{amssymb}
\usepackage{amsthm}
\usepackage{graphicx}
\usepackage{hyperref}
\usepackage{pgfplots}
\usepackage{tikz}
\usepackage{tikz-3dplot}
\usepackage{pgfplots}
\usetikzlibrary{arrows.meta}
\usepackage{fancyhdr}
\usepackage{multirow, multicol}
\usepackage{enumitem}
\usepackage{setspace}

\usepackage{sectsty}
\sectionfont{\large}
\subsectionfont{\normalsize}
\subsubsectionfont{\normalsize	}
\newcommand\R{\mathbb{R}}

\begin{document}
\baselineskip 12pt

\begin{center}
\textbf{\large Principal Component Analysis and Hidden Markov Model for Forecasting Stock Returns} \\
\vspace{1.5cc}
{Eugene W. Park}\\

\vspace{0.3 cm}

{\small A thesis submitted in partial fulfillment\\
of the requirements for the degree of\\
Master of Science \\
Courant Institute of Mathematical Sciences\\
New York University\\
May, 2023}
 \end{center}
\vspace{1.5cc}
\begin{abstract}
  \noindent This paper presents a method for predicting stock returns using principal component analysis (PCA) and the hidden Markov model (HMM), and tests the results of trading stocks based on this approach. Principal component analysis is applied to the covariance matrix of stock returns for companies listed in the S\&P 500 index, and interpreting principal components as factor returns, we apply the HMM model on them. Then we use the transition probability matrix and state conditional means to forecast the factors returns. Reverting the factor returns forecasts to stock returns using eigenvectors, we obtain forecasts for the stock returns. We find that, with the right hyperparameters, our model yields a strategy that outperforms the buy-and-hold strategy in terms of the annualized Sharpe ratio.  

\vspace{0.95cc}
\parbox{28cc}{{\it Keywords}: Principal component analysis, factor model, hidden Markov model, stock market, forecasting }
\end{abstract}
%%%%%%%%%%%%%%%%% INTRO %%%%%%%%%%%%%%%%%%%%%%%%%%%%%%%
\section{Introduction} \label{form}

Stock market forecasting has been a prolonged practice of interest for people in various fields of discipline, and thus, various approaches have been used to tackle the problem from classical time series analysis and using factor models to high frequency trading and using deep learning techniques. 

A widely used method to analyze time series data, the hidden Markov model has been a popular method to analyze financial markets. \cite{Nyugen18}, \cite{Dhingra12}, \cite{Hassan05}, and \cite{Kavitha13} have used the HMM on its own to make predictions of stock prices, while others have combined the HMM with other methods such as the long short term memory model \cite{Huo12} and fuzzy logic \cite{Hassan09}. The usage of the hidden Markov model on stock price prediction is more extensive than what has been cited, but within the author's knowledge, there seems to be more attention on using HMM as a preliminary step than combining techniques that refine the input for the HMM. 

In this paper, we present a method that applies principal component analysis (PCA) as a form of a factor model to preprocess our data and uses the HMM on the preprocessed data to forecast of stock returns. Then, we test the accuracy of this model by computing the directional accuracy of the forecasts, and evaluate the long-term, risk-adjusted returns of various trading strategies based on this model.

The rest of this paper is organized as follows. In section 2, we give a brief overview of the PCA as a factor model, and in section 3, briefly introduce the HMM. Then, we explain our model in detail in section 4, and discuss the implementation and results of our model as trading strategies in section 5. We conclude the paper in section 6 by mentioning some shortcomings of our paper and providing ways for improvement.
%%%%%%%%%%%%%%%%%%% OVERVIEW %%%%%%%%%%%%%%%%%%%%%%%%%%%%%
\section{Brief Introduction: Principal Component Analysis as Factor Models }
\subsection{Principal Component Analysis}
Given a dataset $X \in \mathbb{R}^{t \times n}$ where the rows of $X$ are de-meaned samples of the random vector $\Tilde{X} \in \R^n$, principal component analysis is performed as follows: 
\begin{itemize}
    \item Compute the covariance matrix of $X$: 
\begin{equation}\label{equ:covaraince matrix}
    M = X^TX \in \mathbb{R}^{n \times n}
\end{equation} 
\item Compute the eigendecomposition of $M$, which exists by the spectral theorem for symmetric matrices, such that 
$$ M = EGE^T$$
where $E \in \mathbb{R}^{n \times n}$ is the column matrix of n orthonormal eigenvectors $u_1, \dots, u_n \in \R^n$, and $G \in \mathbb{R}^{n \times n}$ is a diagonal matrix of eigenvalues $\lambda_1, \dots, \lambda_n$. 

Since we are allowed to reorder the eigenvalues in descending order with the eigenvectors ordered accordingly, we may assume that 
$$ \lambda_1 > \dots > \lambda_n$$
Each $\lambda_i$ denotes the amount of variance explained in $M$, and $u_i$ are the corresponding directions called the "principal directions." 
\end{itemize}
Note that $u_1$ is the direction of the highest variance in $M$ and $u_n$ is the direction of the lowest variance. 

The "principal components" are then the dataset $X$ in the directions of the principal direction $u_i$:
$$ w_i = \begin{bmatrix}
    w_{i,1} \\ \vdots \\ w_{i,n}
\end{bmatrix}  
= Xu_i$$
%%%%%%%%%%%%%%%%%%%%%%%%%%%%%%%%%%%%%%%%%%%%%%%%%%%%%%%
\subsection{Factor Models }

The factor model, introduced in the Arbitrage Pricing Theory (APT) in \cite{Ross76}, states that there exist explanatory variables (called "factors") that explain the systematic behavior of asset returns. Hence, 
\begin{equation} \label{equ:factor model}
    \textbf{r} = B\textbf{f} + \epsilon
\end{equation}
where 
\begin{itemize}
    \item $\textbf{r} \in \R^{n \times 1}$ is a random vector of returns for n assets,
    \item $\textbf{f} \in \R^{k \times 1}$ is a random vector of returns for the k factors with $\mathbb{E}(\textbf{f}) = \mu_f$ and $\sigma^2(\textbf{f}) = F$,
    \item $B \in \R^{n \times k}$ consists of columns representing factor loadings for the k factors,
    \item $\epsilon \in \R^{n \times 1} $ is random noise for which we assume $\mathbb{E}(\epsilon) = 0$, $\sigma^2(\epsilon) = D$, a diagonal matrix.
\end{itemize}
From , \ref{equ:factor model} we get 
\begin{equation*}
    \mathbb{E}(\textbf{r}) = B\mu_{\textbf{f}}
\end{equation*}
\begin{equation}
    \sigma^2(\textbf{r}) = \Sigma_{\textbf{r}} = BFB^T + D
\end{equation}
where $\Sigma_{\textbf{r}} = \mathbb{E}(\textbf{r}^T\textbf{r}) \in \R^{n \times n}$ is the covariance matrix of $\textbf{r}$. 
%%%%%%%%%%%%%%%%%%%%%%%%%%%%%%%%%%%%%%%%%%%%%%%%
\subsection{Classical Factor Models and PCA Factor Models}

$\footnote{This subsection is largely based on \cite{winston}}$ Suppose we have a dataset $X$ of n asset returns across t time ($ X \in \R^{t \times n}$). The classical, APT-type, factor models uses features of the assets to derive the k factors that explain $X$. Hence, factors are exogenous to $X$. The Fama-French five factor model \cite{fama} and factor-based risk models built by MSCI are some of the popular examples of the classic factor models. 

The PCA factor models differ from the classical ones in that the factors are not discerned exogenously. The factors are data-driven (i.e. inferred from the data matrix $X$) by applying the eigendecompoistion on the covariance matrix in equation \ref{equ:covaraince matrix}: 
$$ M = X^TX = EGE^T $$
The PCA factor model sets the eigenvectors as factor loadings, 
\begin{equation}\label{equ: E = B}
    E = B 
\end{equation}
, and the factor returns as the n principal components
\begin{equation} \label{equ:factor return}
    \textbf{f} = XE = [w_1, \dots w_n]
\end{equation}
Then, we have 
$$ F = \textbf{f}^T\textbf{f} = (XE)^T(XE)$$
$$ \Rightarrow EFE^T = E(XE)^T(XE)E^T = X^TX$$
$$ \Rightarrow F = G$$
Assuming the PCA model to be exact, hence leaving aside $D$ in (3) (i.e. $\epsilon = 0$ in (\ref{equ:factor model})),  
\begin{equation}
    M = X^TX = EGE^T = BFB^T +0
\end{equation}
Thus, the PCA factor model defines n factors that explain the full covariance matrix of our dataset of returns, $X$, where each factor corresponds to $X$ rotated in n independent direction $u_i \in E$, $i = 1,\dots, n$, and the factor loadings are these n independent directions. $\footnote{Note that the independence of eigenvectors come from the spectral theorem of symmetric matrices, and that the independence of $u_i$'s suggests that $f|_k$ consists of independent column vectors}$ The analogue to (\ref{equ:factor model}) is then
\begin{equation} \label{equ: PCA FM}
     X^T = E(XE)^T
\end{equation}
where $X^T \in \R^{n \times t}$ and $(XE)^T \in \R^{n \times t} $ consists of $t$ samples in the columns. 

We note that the although factors derived by PCA are less intuitive, the PCA offers the benefit of discovering factors that not may be discovered exogenously. 
%%%%%%%%%%%%%%%%%%%%%%%%%%%%%%%%%%%%%%%%%%%%%%%%%%%%
\section{Brief Introduction: Hidden Markov Model}
The fundamental argument of the hidden Markov model (HMM) is that underlying the observed sequence of time series data $Y = \{y_1, \dots, y_t\}$, there exists a Markov chain, $Z = \{z_1,\dots,z_t\}$ that generates $Y$ where each $z_i \in S = \{1,...,N\}$, the state space. The underlying Markov chain has an initial distribution $\textbf{L} = \{L_1, \dots, L_n\}$ and a transition probability matrix $\textbf{P} = (p_{ij})_{i,j\in S}$, where $p_{ij}$ denotes the probability of $z_l$ transitioning from state i to j. Finally, assuming our observations to be in state space $O$, the emission probability matrix, $\textbf{R} = (r_{ij})_{i\in S, j \in O}$, denotes the probability of observation given that we are in a certain state: $r_{ij} = \mathbb{P}(Y = j | Z = i)$.

Hence, the parameters that define the HMM are: 
\begin{equation} \label{equ: HMM}
    \Theta = (\textbf{L}, \textbf{P}, \textbf{R}) 
\end{equation}
and we are mainly concerned with the following problems:
\begin{enumerate}
    \item Selecting the best model $\Theta$ given a range of model options, $\Theta_k$ and the sequence of observations $Y$:
    $$ \underset{\Theta_k}{\text{argmax}}\mathbb{P}(Y|\Theta_k) $$
    \item Determining the most probable state sequence $Z$ given $Y$ and $\Theta$:
    $$ \underset{Z}{\text{argmax}}\mathbb{P}(Z|Y,\Theta)$$
    \item Estimating the parameters $\Theta$ given $Y$:
    $$ \underset{\Theta}{\text{argmax}}\mathbb{P}(Y|\Theta)$$
\end{enumerate}
The \textit{forward algorithm} is an algorithm to solve the first problem, \textit{Viterbi algorithm} for the second, and \textit{Baum-Welch Algorithm} or EM (expectation-maximization) algorithm for the last.$\footnote{Refer to \cite{asa} for details of the algorithms}$
%%%%%%%%%%%%%%%%% DATA AND METHOD %%%%%%%%%%%%%%%%%%%%%%%%%%%%%%%
\section{PCA + HMM as a Forecasting Model} 

In this section, we discuss details for how we use PCA with HMM to develop a model that forecasts asset returns. 

Suppose that we are working with a dataset, $X$, of returns for n assets through T time periods such that the columns of $X$ are time series data of return for a particular asset: 
$$ X = (r_{ij})_{i = \{1,\dots,T\}, j = \{1, \dots, n\}}$$
, where $r_{ij}$ is the return of company j at time i. Thus, our goal is to forecast the returns on the n assets for the next period, $\hat{X}_{T+1,n = \{1, \dots, n\}}$. Furthermore, since the directional accuracy of our forecasts is crucial for our trading strategies, we are, in fact, primarily concerned with:  
\begin{equation}\label{goal:sign} \text{sign}(\hat{X}_{T+1,n = \{1, \dots, n\}})\end{equation}
\subsection{Implementing PCA}
PCA only requires our dataset to be de-meaned, but we normalize $X$ for each column and denote it as $Y$:
\begin{equation}\label{norm}
    Y = \frac{X-\mu_x}{\sigma_X}
\end{equation}
We apply PCA by computing the eigendecomposition of the covariance matrix of $Y$ such that the eigenvalues are in decreasing order and the eigenvectors are ordered accordingly:
$$ H  = Y^TY = EGE^T, \ H,E,G \in \R^{n \times n}$$
$$ \text{diag}G = \{ \lambda_1, \dots, \lambda_n \}, \ \lambda_1 > \dots>\lambda_n $$
Then, as in (\ref{equ: PCA FM}), we have a full PCA factor model: 
$$ Y^T = E(YE)^T$$
We assume that the covariance matrix $H$, or equivalently $Y$, contains some noise. Thus, we want to de-noise our dataset before training the HMM. 

Recall that the eigenvalues suggest the percentage of variance (i.e. information) of $H$ explained by rotating it in the direction of the corresponding eigenvectors. Suppose the amount of noise in the covariance matrix is $p\%$. Then, we take the first $k$ eigenvectors that explain at least $(1-p)\%$ of variance such that 
$$ \lambda_1 + \dots + \lambda_k \geq 1-p$$
$$\lambda_1 + \dots + \lambda_{k-1} < 1-p$$
and compute the corresponding principal components, ($w_1,  \dots, w_k$), which represent k of the n factor returns as in (\ref{equ:factor return}).

Then, we extract $\approx p\%$ $\footnote{$\approx p\%$ is due to the fact that the k eigenvectors does not necessarily, in fact, most often does not, explain exactly $(1-p\%)$ of the matrix.}$ of noise from $H$ by restricting our set of eigenvectors, $E$, to the first k eigenvectors
$$E|_k = \{u_1, \dots, u_k\} \in \R^{n \times k} $$
and computing k factor returns,
\begin{equation}\label{preprocess data} f|_k = \{w_1, \dots, w_k\} = YE|_k = 
\begin{bmatrix}
    y_{1}\cdot u_1 & \dots & y_1\cdot u_k \\
    y_{2}\cdot u_1 & \dots & y_2\cdot u_k \\
    \vdots & \vdots & \vdots \\
    y_T\cdot u_1 & \dots & y_T\cdot u_k 
\end{bmatrix}
\in \R^{T \times k} \end{equation}
, where $y_i \in \R^k$ is the returns for the k factors in time $i$. Note that $f|_k$ consists of columns that represent $Y$ in k independent directions with the most amount of information (i.e. $f|_k$ consists of independent column vectors).

Assuming the full PCA factor model to contain some noise, we are essentially assuming that not all of the principal component factors defined by PCA are significant factors; factors that explain small portions of variance of the covaraince matrix are just noise. Hence, we have:
\begin{equation}\label{equ: restricted y} Y^T = E(YE)^T  = E|_k(YE|_k)^T + \epsilon_k \end{equation}
where $\epsilon_k$ is the error term that contains $\approx p\%$ of information in $H$. 

\subsection{HMM on Factor Returns}
Now, we extract the noise from $Y$ and apply the HMM to $f|_k$, the time series of $k$ factor returns, to obtain a one-step-ahead forecast for these factor returns. Then, we revert the forecast of the factors back to assets which will then be our forecast for the assets. 

Before, we begin training the model, we assume the emission probability in each state in the state space, $S$, to have have a Gaussian distribution. Hence, the parameters of our model are
$$ \Theta = (\textbf{L}, \textbf{P}, \textbf{R})$$
\begin{equation} \label{equ: gaussian hmm}
\Rightarrow \Theta = (\textbf{L}, \textbf{P}, \mu_i, \sigma^2_i), i = 1, \dots, N 
\end{equation},
where N is the number of states in the state space $S$, $\mu_i$ is the mean of the observation sequence in state i, and $\sigma^2_i$ is the variance of observation sequence in state i. 
\subsubsection{Training \& Model Selection}
To determine the number of state space, $N$, we train the Gaussian HMM (\ref{equ: gaussian hmm}) for each $j$ in [2,3,4,5,6,7,8], denoted as $\Theta_j$, compute the log likelihood, $\ln(\mathbb{P}(Y|\Theta_j))$, using the forward algorithm, and compute Akaike information criterion (AIC)\cite{aic} for each model. 
$$ \text{AIC} = -2\ln(\ln(\mathbb{P}(Y|\Theta_j)))+2j$$
Then, we choose our model to be $\Theta_j$, which yields the lowest AIC.

In the training step of each $\Theta_j$, we use the Baum-Welch algorithm to estimate the parameters, and thus, choose the initial conditions for our parameters as follows:  
\begin{equation}\label{init_init} \textbf{L}_0 = 1/j \end{equation}
\begin{equation}\label{init_trans} \textbf{P}_0 = 1/j^2\end{equation}
\begin{equation}\label{init_mu} \mu_{i,0} = \mathbb{E}(o)_i\end{equation}
\begin{equation}\label{init_sig} \sigma^2_{i,0} = \text{var}(o)_i\end{equation}
, where $o$ is the observation sequence, $o = \{o_1, \dots, o_T\}$. Here, we are assuming that it is equally likely for the underlying Markov chain to start in any state (\ref{init_init}) and that it is equally likely for the Markov chain to transit from one state to another (\ref{init_trans}). (\ref{init_mu}) suggests that we assume the $\mu_{i}$ to be the sample mean of our observations for all states $i = 1,\dots, j$ (i.e. same initial mean for all states), and (\ref{init_sig}) suggests that we assume $\sigma^2_{i}$ to be the sample variance of our observations for all states $i = 1,\dots, j$ (i.e. same initial variance for all states). 

Note that since $f|_k$ consists of k time series of factor returns, our observation sequnce is k-dimensional.
$$o = \{o_1, \dots, o_T\}, \text{where } o_t = \{o_t^1, \dots, o_t^k\}$$
Hence, $\sigma^2_{i}$ is a $k \times k$ covariance matrix. Since, the k factor returns are independent, however, we may assume the covariance matrix to be diagonal. Thus, in (\ref{init_sig}), $\sigma^2_{i,0} = \text{var}(o)_i$ is a $k \times k$ diagonal covariance matrix, for state $i$, with sample variances of $o = \{o^1, \dots, o^k\}$ in the diagonal. 

\subsubsection{Forecasting}
Now, once we have found and trained our model, which we will denote again as $\Theta$, we use the estimated parameters to forecast the returns of the k factors in the next time period, $(\hat{f|_k})_{T+1}$, as follows: 
\begin{enumerate}
    \item We use the \textit{Viterbi algorithm} to obtain the sequence of the underlying Markov chain: 
    $$ Z = \{z_1, \dots, z_T\}, \ z_i \in S = \{1, \dots, N\}$$
    , where $z_T = i$ states that the underlying Markov chain is currently in state $i$.
    \item Then, we use the transition probability matrix, $\hat{\textbf{P}}$, and the state conditional mean, $\hat{\mu}_i$, to obtain an estimate for $(f|_k)_{T+1}$:
    \begin{equation}\label{factor forecast}
        (\hat{f}|_k)_{T+1} = \displaystyle \sum_{j \in S}^N P_{ij}\hat{\mu}_j
    \end{equation}
    Since $\hat{\mu}_j \in \R^k$, for all $j = \{1,\dots,N\}$, $(\hat{f}|_k)_{T+1} \in \R^k$. 
\end{enumerate}
Thus, our forecast of the k factor returns in the next period depends on 1) how likely it is to transition from state $i$ to $j$, and 2) estimated state conditional means of the k factors. 

Although we may trade factors, especially if it were defined exogenously as in the classical factor models, it is hard to trade endogenous factors defined by the PCA. Hence, we revert the forecasts for the factor returns to forecasts for the $n$ assets. I.e. we want: 
$$(\hat{f}|_k)_{T+1} \rightarrow \hat{Y}_{T+1}$$
Recall from (\ref{equ: restricted y}) that 
$$ Y^T = E(YE)^T = E(f)^T$$
where $f = \{w_1,\dots, w_n\}$. Since we only have an estimate of $f|_k$ for time $T+1$, we first estimate $f_{k+1},\dots,f_n$ for $T+1$. Recall that the best estimate is its mean. Thus, we have, for $j = k+1,\dots,n$,
\begin{equation} \label{equ: est f_} (\hat{f}_{j})_{T+1} = \mathbb{E}((f_j)_{T+1}) =\frac{1}{T}\displaystyle \sum_{i = 0}^T(Y_i \cdot u_j) = \frac{1}{T}\displaystyle \sum_{i = 0}^T(Y_i) \cdot u_j= 0\end{equation}
where $Y_i \in \R^n$ is the i-th row of $Y$, and the last equality follows from $Y$ being normalized with row mean equal to 0. Equivalently, 
$$ \mathbb{E}((\hat{\epsilon_k})_{T+1}) = \mathbb{E}(\hat{Y}_{T+1}^T - E|_k(\hat{Y}_{T+1}E|_k)^T) = \frac{1}{T}\displaystyle \sum_{h = 0}^TY_h^T - E|_k(\frac{1}{T}\displaystyle \sum_{h = 0}^TY_h^T)E|_k)^T =0  $$
where $Y_h^T$ are the h-th column of $Y^T$.

Now, we have 
$$ \hat{f}_{T+1} = \{ (\hat{f}|_k)_{T+1}, 0 , \dots, 0 \} $$
and from (\ref{equ: restricted y}), we have
$$ \hat{Y}^T_{T+1} = E(\hat{f}_{T+1})^T = E|_k(\hat{f}|_k)^T_{T+1} $$
$$ \Rightarrow \hat{Y}_{T+1} = (\hat{f}|_k)_{T+1}E|_k^T $$
where $\hat{Y}_{T+1} \in \R^{1 \times n}$, $(\hat{f}|_k)_{T+1} \in \R^{1\times k}$, and $E|_k^T \in \R^{k \times n}$.
 Thus, we obtain forecasts for the $n$ assets by multiplying the transpose of the k eigenvectors to the forecasts for the k factor returns. 

 Finally, since $Y$ is the normalized version of $X$ (\ref{norm}), 
 \begin{equation}
     \hat{X}_{T+1} = \sigma_X(\hat{y}_{T+1}) + \mu_X
 \end{equation}
%%%%%%%%%%%%%%%%%%%%%%%%%%%%%%%%%%%%%%%%%%%%%%%%
\section{Implementation}
We use data for the weekly returns of companies listed in the S\&P500 to train our model.$\footnote{Data was obtained from finance.yahoo.com}$$\footnote{List obtained from https://en.wikipedia.org/wiki/List\_of\_S\%26P\_500\_companies on  April 28, 2023}$ The returns are calculated using the closing stock prices, and each return sequence comprise the columns of our dataset, denoted $X$. Hence, we have $X \in \R^{T x n}$, where $T$ is the last period of the time span of our data and $n$ is the number of assets. 

We note that, as companies are newly listed and delisted from the S\&P500 index, some companies that are listed in the index at the time of data retrieval may not have been listed previously, and some may have been delisted during the time span of our data. Thus, we are subject to \textit{survivor bias}, but minimize this effect by ensuring that we have at least 400 data points of company returns for each time period. With at least 80\% of the index, we conjecture that the several companies we fail to capture only explains a minimal amount of information/variance in the covariance matrix, $X^TX$, which we implement the PCA on. In other words, we suspect that most of the information from the companies we fail to capture have already been captured by the eigenvectors corresponding to large eigenvalues, and that the additional information from those companies constructs eigenvectors with very small eigenvalues. Thus, we suspect that they are eventually extracted during the process of denoising our covariance matrix via PCA. Because this is only a conjecture, however, we do not know for sure how much we are affected by survivor bias, and thus, leaving room for improvement upon this paper.  

\subsection{Model Training}
Ensuring our dataset to include returns data for at least 400 companies listed in the S\&P500 index for each $t$, we start our data from July 23, 2004. Using 10 years' worth of weekly data, we train our PCA+HMM model and forecast the weekly returns for the next period. Then, rolling the 10 years window by one week$\footnote{The exact time samples (rows of our dataset) for each window may slightly differ due to different numbers of holidays and trading days every year. Moreover, due to the entering and exiting of companies from the index since the beginning of our dataset, the number of stocks (columns of our dataset) may also differ.}$, we forecast returns for a total of 100 weeks (the week for which we make our last forecast is July 8, 2016, thus our data spans from July 23, 2004 to July 8,2016)$\footnote{Additional data is available, but we limit ourselves to 100 forecasts for computational convenience.}$. Note that, since we train the PCA+HMM for each time window, the total number of PCA factors, the total number of states in the HMM, and the model parameters may vary over time.  

Since PCA is essentially an eigendecomposition, We use the \textit{linalg} module in the \textit{numpy} library in Python to implement PCA. To train the HMM model, we use the \textit{gaussianhmm} module in the \textit{hmmlearn} Python library.

Recall from part 4.1 that we have to specify the hyperparameter, $p$, that specifies the minimum percentage of noise to be extracted during the dimension reduction process. We train our model with four choices of $p$: 
$$ p = \{45\%, 30\%, 15\%, 10\% \} $$
which corresponds to keeping no less than  55\%, 70\%, 85\%, and 90\% of the information in $Y$, our normalzied return matrix.

\subsubsection{Model Implications}
Before, we test trading strategies based on our model, we make several observations of our trained model. 

\begin{figure}[h]
  \centering
  \subfigure[P = 45\%]{\includegraphics[width=0.44\linewidth]{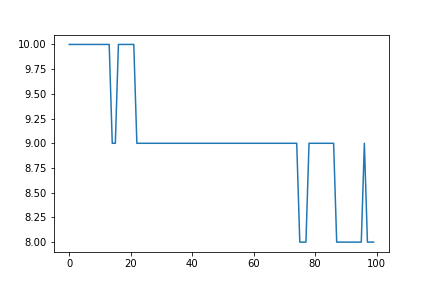}}
  \hfil
  \subfigure[P = 30\%]{\includegraphics[width=0.44\linewidth]{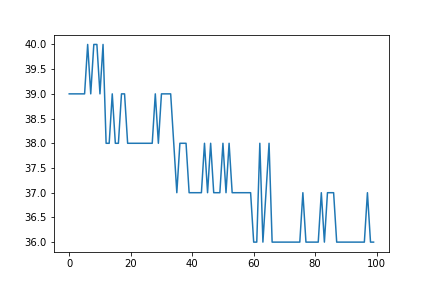}}
    \hfil
  \subfigure[P = 15\%]{\includegraphics[width=0.44\linewidth]{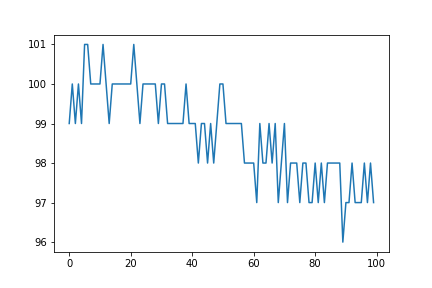}}
    \hfil
  \subfigure[P = 10\%]{\includegraphics[width=0.44\linewidth]{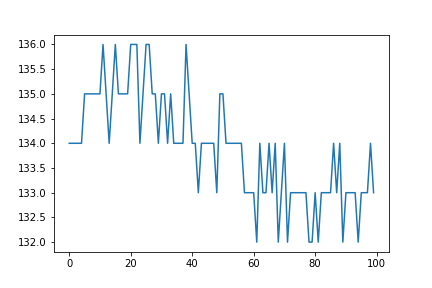}}
  \caption{Number of PCA Factors}
  \label{fig:pca_num}
\end{figure}

Figure \ref{fig:pca_num} shows the total number of PCA factors kept in the model throughout time for the different choices of $p$. Notice that as we keep more noise in our model, the number of PCA factors kept not only increases, but also changes more frequently. Moreover, there is also a more gradual change in the total number of PCA factors in the market, such as from the first 20 periods to the period between the 20th and 60th week in figures \ref{fig:pca_num}.a and \ref{fig:pca_num}.b. There is also a gradual shift in \ref{fig:pca_num}.c and \ref{fig:pca_num}.d from the first 50 weeks to the next. This suggests that there may exist short-term PCA factors and longer-term PCA factors.

Now, we observe the state-conditional means and variances calibrated from the HMM model. Recall that the HMM calibrates the means and variances for $K$ PCA factors for each state, and since we run the model for 100 periods, we have a series of state-conditional means and variances as follows:
$$ \mu = \{\mu_{1}, \dots, \mu_{100}\}$$
$$ \sigma^2 = \{\sigma_{1}^2, \dots, \sigma_{100}^2\}$$
where $\mu_{t}, \sigma_{t}^2 \in \R^{N_t \times k_t}$. Note that $N_t$ and $k_t$ has a time subscript since the total number of states and PCA factors vary. 

\begin{figure}[h]
  \centering
  \subfigure[P = 45\%]{\includegraphics[width=0.44\linewidth]{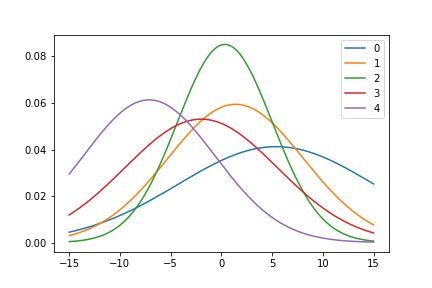}}
  \hfil
  \subfigure[P = 30\%]{\includegraphics[width=0.44\linewidth]{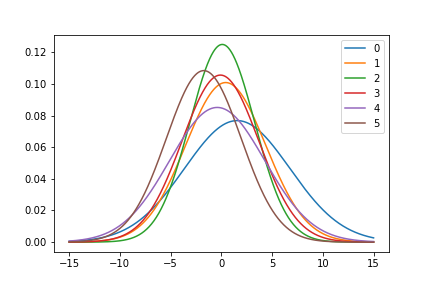}}
    \hfil
  \subfigure[P = 15\%]{\includegraphics[width=0.44\linewidth]{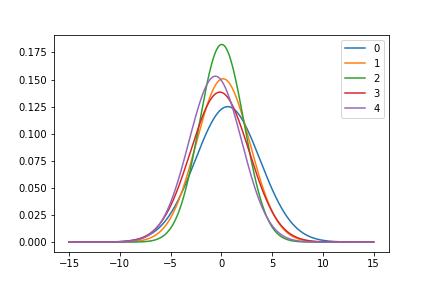}}
    \hfil
  \subfigure[P = 10\%]{\includegraphics[width=0.44\linewidth]{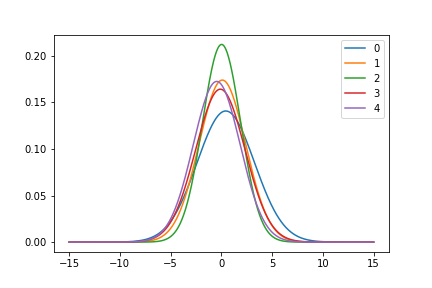}}
  \caption{State-Conditional Emission Probability Distributions}
  \small\textbf{Note:}x-axis in percentages. 3 refers to 300\%. 
  \label{fig:prob_dist}
\end{figure}

At each time, $t$, we average $\mu_{t}, \sigma_{t}^2$ along the k factors to get a time series of state-conditional means and variances (of returns)
$$ \Tilde{\mu} = \{\Tilde{\mu}_{1}, \dots, \Tilde{\mu}_{100}\} $$
$$ \Tilde{\sigma}^2 = \{\Tilde{\sigma}_{1}^2, \dots, \Tilde{\sigma}_{100}^2\}$$
where $\Tilde{\mu}_{t}, \Tilde{\sigma}_{t}^2 \in \R^{N_t}$. 
Moreover, since the total number of states are changing, we sort $\Tilde{\mu}_{t}$ in decreasing order with $\Tilde{\sigma}_{t}^2$ having the corresponding order. Thus, we disregard the actual state numbers, such as rather we are in state 1 or state 2, and put meaning on the states by ranking them according to the calibrated state-conditional means averaged across the $k_t$ PCA factors. For each $p$, we find the minimum number states that existence throughout time and visualize the emission probability distributions in figure \ref{fig:prob_dist}.

From figure \ref{fig:prob_dist}, we can see that more noise being kept in our factor returns yields probability distributions that are less spread across states, suggesting that the underlying states lose significance and the HMM model becomes less useful. Furthermore, figure \ref{fig:prob_dist}.a, the model with $45\%$ of noise extracted from the normalized return dataset,$Y$, is intuitively interpretable. State 4 to seems to be "bear market" state, while state 0 to can be viewed as the "bull market" state, and states 1,2, and 3 are the relatively stable states of the market. We can also view state 4 as where investors have excessively fearful, state 0 as the state where they are excessively optimistic, states 1 and 3 as states where they are moderately optimistic and fearful, and state 2 as where they neutral about the market. Thus, models with more spread-out probability distributions provide more reliable and interpretable information about the market which may be useful for task such as risk-management. 

\subsection{Trading Strategies} 
Now we construct and test simple trading strategies based on our model. 
\begin{enumerate}
    \item Our first trading strategy is solely based on the forecasts given by \ref{asset forecasts}. We short the assets for which $\hat{X}_{T+1}$ predicts a negative return and long the ones for which $\hat{X}_{T+1}$ predicts a positive return. We weight the $n$ stocks equally (i.e. 1/n) and trade at the closing time of the stock market at the last trading day of every week. 
    \item The second strategy depends on the forecasts of the normalized asset returns, $\hat{y}_{T+1}$. Since $$Y = \frac{X-\mu_X}{\sigma_X}$$, we hypothesize that there is predictive trading signal in not the return itself, but the excess return over the average.
\end{enumerate}
When testing our trading strategies, we assume the following:
    \begin{enumerate}
        \item we are able to trade at exactly the closing time 
        \item we can exactly match the $n$ assets with equal weight in our portfolio
        \item and we do not incur any trading costs.
    \end{enumerate}

\subsection{Results}
To evaluate the performance of our trading strategies, we compare our trading strategies with the simple buy-and-hold strategy of all companies in the index. As measurements of performance, we compute: 
\begin{enumerate}
    \item the "winning probability" of our forecasts (i.e. probability of correctly forecasting the direction of the returns in the next period). 
    \item the annualized Sharpe ratio: 
    \begin{equation}
        S = \frac{r - rf}{\sigma}
    \end{equation}
    where $r$ is the asset return, $rf$ is the return of the risk-free rate, and $\sigma$ is the standard deviation of asset. We will assume the risk-free rate to be zero. 
\end{enumerate}

\begin{table}[h]
  \centering
  \begin{threeparttable}
    \captionsetup{width=0.8\linewidth}
    \caption{Winning Probability}\label{tab:wp}
    \begin{tabular}{ccc}
      \toprule
      p\% & Strategy 1 & Strategy 2 \\
      \midrule
      45\% & 0.532 & 0.490 \\
        30\% & 0.543 & 0.504 \\
      15\% & 0.538 & 0.511 \\
      5\%  & 0.532 & 0.490 \\
      \bottomrule
    \end{tabular}
    \begin{tablenotes}[flushleft]
      \footnotesize
      \item \textit{Note:} Rounded to the nearest thousandth.
    \end{tablenotes}
  \end{threeparttable}
\end{table}

The winning probabilities of the two trading strategies are shown in table \ref{tab:wp}. We observe that strategy 1 clearly outperforms strategy 2 in forecasting the direction of the returns. Yet, strategy 1 yield just above 50\%, suggesting that significant profit can be made only in a casino-style type of trading where the model trades constantly without going bankrupt. Hence, for this strategy to be profitable over the long-run, additional features must be included such as the trade size. 

\begin{figure}
  \centering
  \includegraphics[width=0.6\textwidth]{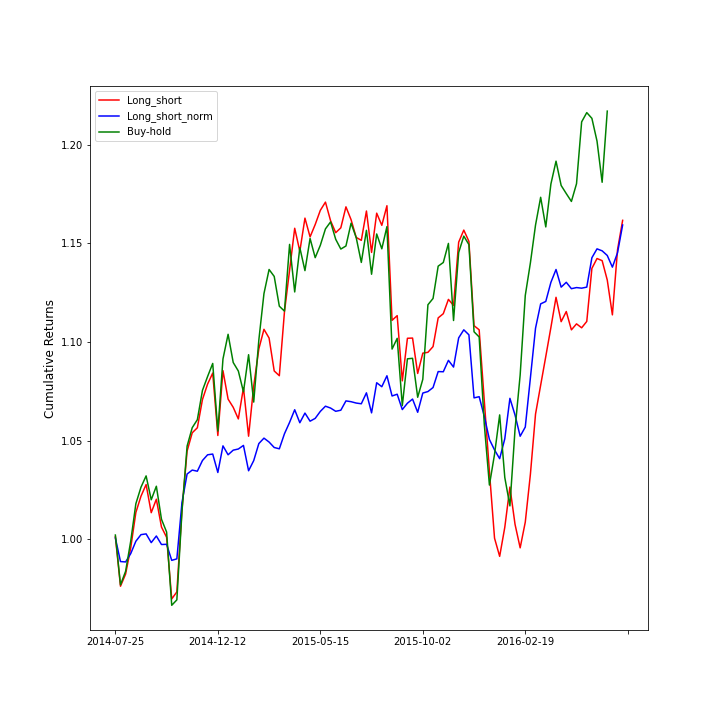}
  \caption{Return Sequences for Strategies 1 and 2 with p = 15\%}
  \label{fig:plot85}
\end{figure}

Table \ref{tab:sr} shows the Sharpe ratios of the trading strategies, and we observe that the strategy with the highest Sharpe ratio is strategy 2 with $p = 15\%$. This outperforms the buy-\&-hold strategy by more than 50\%. From figure \ref{fig:plot85}, we observe that the strategy yields lower cumulative return than the buy-\&-hold, but is more stable; it rides out two ocassions of downturns, during late-2014 and early-2016, without much losses, contrary to the other strategies. It also has a winning probability of 51\%, and thus, seems to be useful in terms of risk management and as a stable component within a portfolio. 

We also note that the Sharpre ratios depend heavily on the hyperparameter $p$. The difference between the highest and lowest ratios of strategy is as high as 0.91. For strategy 1, there seems to be a trend where the ratio falls as $p$ decreases to $30\%$, but increases again as more noise is extracted. This suggests the significance of finding the right hyperparameter $p$, which could yield significantly higher Sharpe ratios than the ones presented here. 

\begin{table}[h]
  \centering
  \begin{threeparttable}
    \captionsetup{width=0.8\linewidth}
    \caption{Annualized Sharpe Ratio}\label{tab:sr}
    \begin{tabular}{ccc}
      \toprule
      p\% & Strategy 1 & Strategy 2 \\
      \midrule
      45\% & 0.688 & 0.45  \\
      30\% & 0.581 & 0.581   \\
      15\% & 0.703 & 1.36  \\
      10\%  & 0.877 & 0.726  \\
      \midrule
      Buy-\&-Hold & 0.828 \\
      \bottomrule
    \end{tabular}
    \begin{tablenotes}[flushleft]
      \footnotesize
      \item \textit{Note:} Values are percentages. 
    \end{tablenotes}
  \end{threeparttable}
\end{table}

\begin{figure}
  \centering
  \includegraphics[width=0.6\textwidth]{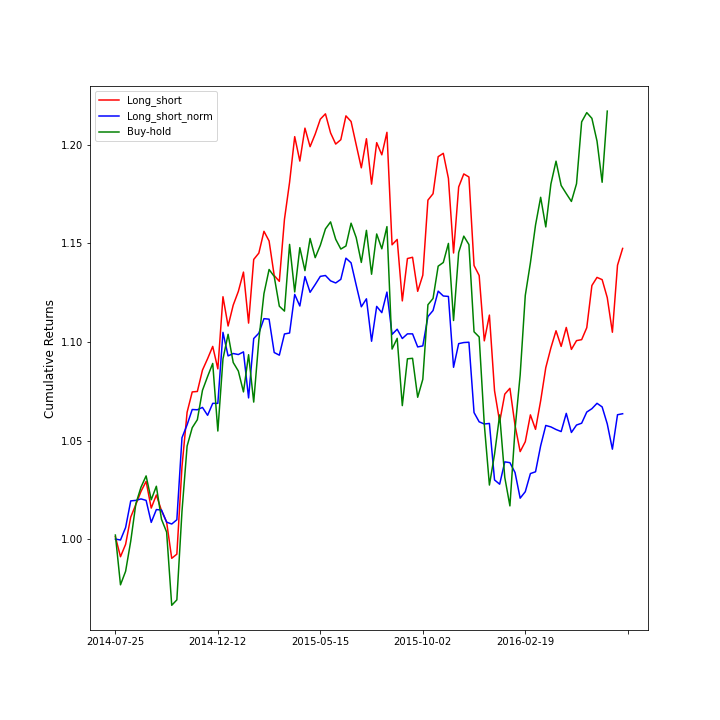}
  \caption{Return Sequences for Strategies 1 and 2 with p = 45\%}
  \label{fig:plot55}
\end{figure}
\begin{figure}
  \centering
  \includegraphics[width=0.6\textwidth]{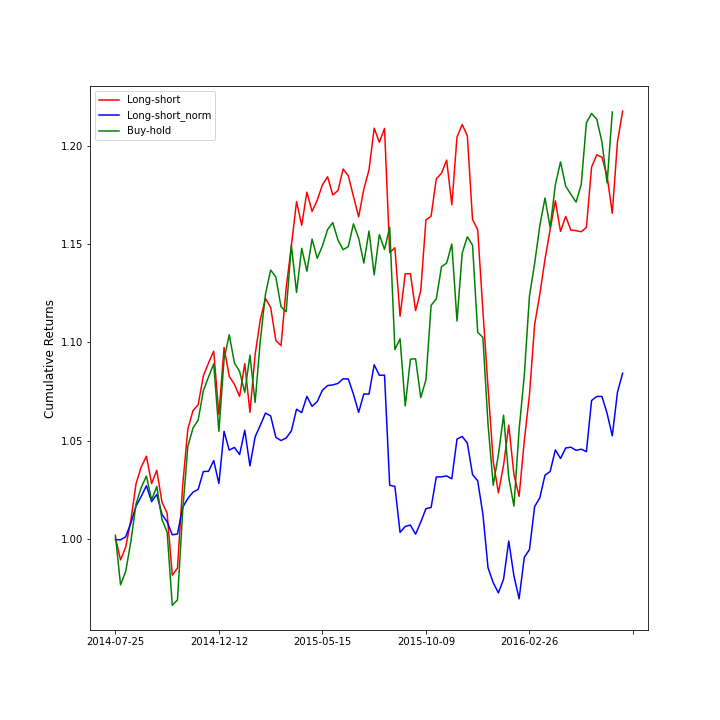}
  \caption{Return Sequences for Strategies 1 and 2 with p = 30\%}
  \label{fig:plot70}
\end{figure}
\begin{figure}
  \centering
  \includegraphics[width=0.6\textwidth]{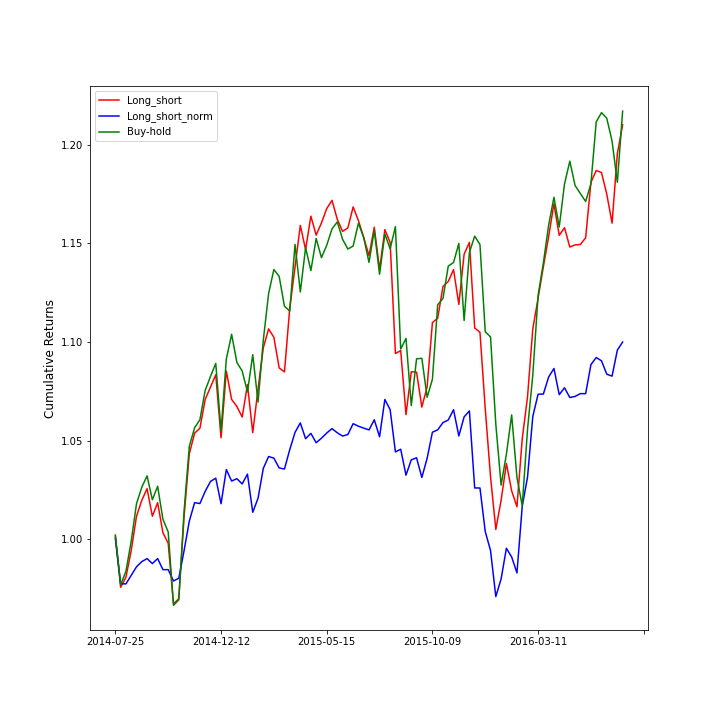}
  \caption{Return Sequences for Strategies 1 and 2 with p = 10\%}
  \label{fig:plot90}
\end{figure}

\section{Conclusion}
We have shown that with the right hyperparameter $p$, the PCA + HMM model provides an intuitive understanding of the market through the estimated emission probability distributions of the states, and yields model-based trading strategies with forecast accuracy slightly above 50\%. We have also seen that these strategies have the potential to outperform the market by finding the right $p$. This work suggests that future research into finding the optimal p will be very useful. Moreover, we have made several assumptions which such as zero transaction costs and that we are able to trade at exactly the closing prices at the closing time. These assumptions are unrealistic and fails in practice. Hence, our paper could be improved by evaluating the performance under realistic conditions, such as existence of slippage, transactions costs, and possibly price impact as well (when we assume large trade size). 
\vfill
{\small \textbf{Acknowledgement:} We thank professors Kenneth Winston, Petter Kolm, and Jonathan Goodman for their valuable comments and suggestions throughout the progress of this work.} 

\newpage

\section*{Appendix}
\textit{Appendix A.1. Proof of Forecast Equation \ref{factor forecast}:}\\
\indent Denote $r_t$, return at time t, as the rows of $f_t$.
\begin{flalign}
    (\hat{f}|_k)_{T+1} & = \mathbb{E}(r_{T+1}|Z_T = i) \nonumber\\
    & = \displaystyle \sum_{j \in S}^N\mathbb{E}(r_{T+1}, Z_{T+1} = j|Z_T = i) \nonumber\\
    & =\displaystyle \sum_{j \in S}^N\mathbb{E}(r_{T+1}| Z_{T+1} = j) \mathbb{P}(Z_{T+1} = j | Z_T = i)\nonumber \\
    & = \displaystyle \sum_{j \in S}^N P_{ij}\hat{\mu}_j\nonumber
\end{flalign}

\end{document}